\documentstyle[12pt]{article}
\begin{document}
\def\b{\bar}
\def\C{{\Cal C}}
\def\d{\partial}
\def\D{\Delta}
\def\cD{\Cal{D}}
\def\e{\epsilon}
\def\E{{\Cal E}}
\def\f{\varphi}
\def\F{\Cal{F}}
\def\g{\gamma}
\def\G{\Gamma}
\def\l{\lambda}
\def\L{\Lambda}
\def\M{{\Cal M}}
\def\m{\mu}
\def\n{\nu}
\def\O{\Omega}
\def\p{\psi}
\def\q{\b q}
\def\r{\rho}
\def\t{\tau}
\def\x{\phi}
\def\X{\~\xi}
\def\~{\tilde}
\def\h{\eta}
\def\z{\zeta}
\def\Z{{\b\zeta}}
\def\Y{{\b Y}}
\def\cZ{{\b Z}}
\def\`{\dot}
\hfill{hep-th/9503094}
\par
\bigskip
\hfill{Dedicated to \ the \ memory of}
\par
\hfill{ Professor Dmitri IVANENKO}
\par
\vskip3cm
\centerline{\bf COMPLEX STRING AS SOURCE OF KERR GEOMETRY
\footnote{To be published in the Memorial Issue dedicated to
Professor Dmitri Ivanenko, Problems of Modern Physics, Especial Space
Explorations v.9, 1995}}
\vskip1cm
\par
\centerline{A. Ya. Burinskii}
\centerline{Nuclear Safety Institute of the Academy of Sciences }
\centerline{B. Tulskaya 52  Moscow 113191 Russia,
 e-mail:grg@ibrae.msk.su}
\par
\bigskip
ABSTRACT
\par
\medskip
\begin{quotation}
   The Kerr solution is considered as a soliton-like background for spinning
elementary particles.  Two stringy structures may be found in the Kerr
geometry, one string is real and another one is complex. The main attention
is paid to the complex string, which is connected with the Lind-Newman
representation of the Kerr metric source as an object with a complex
world-line. In a separate note, one new result is announced concerning
the real string: in the dilaton-axion gravity the Kerr singular ring may be
considered as a heterotic string, the field near the Kerr singularity
coincides with the solution obtained by Sen for heterotic string [23].
\end{quotation}
\par
\newpage
1. The Kerr solution is well known as  a  field  of  the
rotating black hole. However, for the case of  a  large  angular
momentum $L$; $\mid a\mid  = L/m \geq  m$  all the horizons of the Kerr
metric are absent and one appears the  naked  ring-like  singularity.
This naked singularity has many unpleasant manifestations and must be
hidden  inside a rotating disk-like source. The Kerr solution  with $\mid
a\mid  \gg  m$   displays some remarkable features  indicating  a relation
to the structure  of   the   spinning   elementary particles.
\par
In the 1969
Carter  [1]  observed,  that  if  three parameters of the Kerr - Newman
metric are adopted to be ($\hbar $=c=1 ) $\quad e^{2}\approx  1/137,\quad
m \approx 10^{-22},\quad a \approx  10^{22},\quad ma=1/2,$
then one obtains  a  model  for  the four  parameters  of  the electron:
charge, mass, spin and  magnetic moment, and the giromagnetic ratio is
automatically the same as that of the Dirac electron.  Some interesting
considerations were given in this connection by Israel [2].
  Then followed a  model of "microgeon" with Kerr metric [3] and
an analogy  of this model  with the string models [4-6].  L\'opez
[7] has given a very interesting model of the Kerr-Newman source, which
generalizes the classical model of the electron for the case of a spinning
particle.
The L\'opez source represents a rigid rotating thin shell in the  form  of
a highly oblate spheroid.
Material of the source  must   have very
exotic properties: null energy density and negative  pressure  .  These
properties were explained on the basis  of the  volume  Casimir effect
in work [6]. The electromagnetic properties of the material  are close to
those of a superconductor [6,7,8], that allows to consider singular ring of
the Kerr source as  a closed vortex string like the
Nielsen-Olesen  and Witten [9] superconducting strings.
\footnote{See also the note added at the end of the paper.}
\par
On the other hand it has been shown recently in works [10,11] that from
the complex point of view the complex source of  Kerr  geometry may be
considered as a complex hyperbolic string with an orbifold-like structure
of world-sheet.
\par
The real Kerr solution possesses a string-like singularity instead of the
point-like  singularity of the Schwarzschild or the Reisner - Nordstr\"om
 solutions.  This singular ring corresponds to
$ \tilde{r} = \sqrt{ (x-x_o)^2+(y-y_o)^2 +(z-z_o)^2}  =0 $
and arises on the real
 slice of space-time  if the point-like source is placed at a complex
 point of space $(x_o (t), y_o (t), z_o (t))$.
\footnote{It was obtained by Appel still 1887 (!) [12] and may be
used to get the Newton or Coulomb analogues of the Kerr solution.}
Thus the world line of the point-like source turns into a two-dimensional
surface or a world-sheet since the source is smeared now over the singular
ring. Correspondingly a path integral for a point-like
  particle $ \int d t \{dX\} e^{i S_{cl}},  $ turns for the Kerr source
into an integral over the world-sheets.
\par
  However from complex-analytical point of view the Kerr source may be
considered as a "particle" propagating along a complex world-line [10,11,13]
parametrized by complex time.
\par
The world-lines with complex time has been used also in some models
of spinning and superparticles. Complex time is very useful also in string
theory by the conformal methods and representation the strings on the
Riemann sphere. The objects described by the
complex world-lines occupy an intermediate position between  particle
and  string. Like the string
they form the two-dimensional surfaces or the world-sheets in the
space-time. Because of the Euclidean signature they
 were called "hyperbolic strings" [14].
It was shown [10] that the complex Kerr source may be considered as
hyperbolic strings and require an orbifold-like structure of
the world-sheet, that induces a related orbifold-like
structure of the Kerr geometry.  Such an
orbifold-like structure was recently suggested for two-dimensional
black holes also by Witten [15] and considered by Horava [16].
\par
\bigskip
2. The  Kerr geometry may be obtained from the complex source by a retarded
time construction with using of the complex light cones
adjoined to the points
of the complex world line. The real slice of the cones
forms the shear free and geodesic null congruence of the Kerr space, which
 plays a fundamental role in structure of Kerr geometry.
So, the Kerr metric may be represented in the Kerr-Schild form [17]
 \begin{equation}
g_{\m\n} =
\h_{\m\n} + 2 h e^3_{\m} e^3_{\n}, \label{1.4} \end{equation}
where $\h$ an auxiliary Minkowski space
$\h = dx^2 + dy^2 = dz^2 -dt^2 ,$ and $h$ a scalar function.
The principal null directions $ e^3 $ are tangent to the null geodesic
and shear free congruence.
 In the null coordinates
\begin{equation}
2^{1\over2}\z = x+iy ,\qquad 2^{1\over2} \Z = x-iy ,\qquad
2^{1\over2}u = z + t ,\qquad 2^{1\over2} v = z - t  \label{1.6}
\end{equation}
the congruence is  defined by the complex function $Y (x) $
\begin{equation}
e^3 = du+ \Y d \z  + Y d \Z - Y \Y d v.
\label{1.7} \end{equation}
\par
 {\bf The Kerr theorem} [17,18,19] gives a general rule to construct
such congruences in twistor terms. In accordance with the Kerr theorem
an arbitrary  geodesic shear free principal null congruence is defined by
function $Y (x)$ which is a solution of the equation
      $$         F  = 0 ,                        $$
where   $F (\l_1,\l_2,Y)$   is an arbitrary analytic function of
projective twistor coordinates
\begin{equation}
 \l_1 = \z - Y v, \qquad \l_2 =u + Y \Z, \qquad Y.
\label{1.19} \end{equation}
The singularities of the  solutions are caustics of the congruence and they
are defined by the system of equations
\begin{equation}
 F=0, \qquad \d_Y F =0.
\label{1.23} \end{equation}
\par
In the Kerr solution the congruence belongs to a class having the
singularities  contained in a bounded region of space.
  In this case the function $F$ is quadratic in $Y$ and the equation $ F=0$
may be easily solved in explicit form [5,10,17,20]. Besides, one can
represent $F$ in a form which shows that the Kerr congruence depends on a
straight complex world line via a retarded - time construction.
We introduce a complex world line  $ x_0^\m (\t)$ parametrized by the
complex time parameter $\t$ (the corresponding null coordinates of this
world line are $ (\z_0, \Z_0, u_0 ,v_0)$), and also introduce an operator
\begin{equation}
 K(\t) = \`x_0^\m(\t) \d_\m. \label{1.27} \end{equation}
In analogy with twistor coordinates
\ref{1.19}
we introduce a special notation for the values of
 the twistor on the points of world-line
\begin{equation}
 \l_1^0 (\t)= \z_0(\t) - Y
v_0(\t), \qquad \l_2^0(\t) =u_0(\t) + Y \Z_0(\t).\label{1.30}
\end{equation}
Then the new form of $F$ is
\begin{equation}
F \equiv (\l_1 - \l_1^0) K \l_2 - (\l_2 -\l_2^0) K \l_1 .\label{1.26}
\end{equation}
  For the straight world line corresponding the Kerr solution
\begin{equation}
 x_0^\m (\t) = x_0^\m (0) +
V_0^\m \t; \qquad V_0^\m = (1,\b V), \label{1.28} \end{equation}
and $K$ is independent
 of $\t$ and is the Killing vector of the Kerr  solution $K=V_0^\m
 \d_\m$.
\par
 To obtain the geometrical properties of this equation we
consider a complex light cone with the vertex at some point $x_0$ of the
complex world line $x_0^\m(\t)$. The solutions of the complex light cone
equation
\begin{equation}
(x_\m - x_{0 \m})(x^\m -x_0^\m) = 0 ,\label{1.33} \end{equation}
split into two families of  null planes:  "left" planes
\begin{equation}
 x_L = x_0(\t) + \alpha e^1 + \beta e^3    , \label{L} \end{equation}
and"right"planes
\begin{equation}
 x_R = x_0(\t) + \alpha e^2 + \beta e^3, \label{R} \end{equation}
where $ \alpha$ and $\beta$ are arbitrary parameters
on these planes.
The null tetrad
$ e_1, e_2, e_3, e_4 $ is given by conditions
\begin{equation} g_{ab}=e_a^\m e_{b\m} = g^{ab}, \label{1.1} \end{equation}
where
\begin{equation} g_{12}= e_1 e_2 = g_{34}= e_3 e_4 = 1, \qquad (others
\quad components \quad g_{ab}) = 0 .
\label{1.2}
\end{equation}
$e^3, e^4$ are real null vectors, $ e^1, e^2 $ are complex conjugates.
\footnote{The null tetrad $e_a^\m$ complete [17] as follows:
\begin{equation}
e^1 = d \z - Y d v;  \qquad
e^2 = d \Z - \Y d v; \qquad
e^4 =  d v - h e^3.
   \label{1.8}\end{equation}
The inverse tetrad has the form
\begin{equation}
 \d_1 = \d_\z
 - \Y \d_u ; \qquad \d_2 =  \d_\Z - Y \d_u ; \\
\d_3 =  \d_u - h \d_4 ;
\qquad \d_4 = \d_v + Y \d_\z + \Y \d_\Z - Y  \Y \d_u .   \label{1.9}
\end{equation}
The p.n. congruence is geodesic and  shear free [17] if $Y,_2 = Y,_4 = 0. $}
Obviously (\ref{L}) and
(\ref{R}) are solutions of (\ref{1.33}) in consequence of
(\ref{1.2}).  The twistor parameters $\l_1$ and $\l_2 $ may
be represented in the form
\begin{equation}
 \l_1 = x^\m e^1_\m , \qquad \l_2 = x^\m
(e^3_\m - \Y e^1_\m). \label{1.34} \end{equation}
  Substitution
(\ref{L} in
(\ref{1.34})
shows that for an arbitrary Y the twistor parameters $ \l_1$ and $\l_2$
are constants on the "left" planes
\begin{equation}
 \l_1 = \l_1^0(\t);\qquad \l_2
= \l_2^0(\t), \label{1.35} \end{equation}
and the equation $F = 0$ is fulfilled. Real cut
of a "left" null plane gives a real null ray which is defined by
the twistor $\l_1^0(\t), \l_2^0(\t), Y$.  Thus the Kerr principal null
congruence arises as the real cut of the family of the "left" null planes
of the complex light cones which vertices lie on the straight complex
world line $x_0(\t)$.
\par
The parameter $\t$ may be defined for each point $ x$ of the
Kerr space-time and plays really the role of a retarded time parameter.
Its value for a given point $x$ may be defined by using the solution $Y(x)
$ and by forming the twistor parameters
(\ref{1.19}) which fix the
"left" null plane
(\ref{1.35}). A point of intersection this plane
with the complex world-line $x_0(\t)$ gives a value of the "left"
retarded time $\t_L$.  Thus $\t_L$ is in fact a complex scalar function on
the space-time $\t_L(x)$.
\par
\medskip
3. This  representation is working also in nonstationary case.
Let $x_0(\t)$ is an arbitrary complex world-line parametrized
by complex time parameter $\t$ (not only straight and so far not only
analytical).  For every point $x$ of the space-time one can consider the
complex light cone and the point $x_0(\t)$ of the intersection of this
light cone  with the complex world line.  One can look for a solution for
parameter $\t$ of the corresponding light cone equation
\begin{equation}
\t = t - r (x,
x_0 (\t)).  \label{K1} \end{equation}
 This is an implicit nonlinear equation concerning
the retarded time coordinate $\t$. Its solution $\t(x)$ is a complex
scalar function on the space-time. One can introduce $K(\t) = \`x_0^\m(\t)
\d_\m$ and from the light cone equation $(x_\m - x_{0 \m})(x^\m -x_0^\m) =
0 $ one  finds that
\begin{equation}
K(\t) \t(x)  = 1,\qquad K \l_1^0 = K \l_1;\qquad
K \l_2^0 = K \l_2 .\label{1.39} \end{equation}
\par
We can find now the conditions on the
function $ F $, which will guarantee that its solution Y (x) gives
a shear free and geodesic congruence.
After quite long calculations we find [18]:
\par
  i) the world-line $ x_0(\t)$ must have an analytical dependence on
 the complex time $ \t,$ or
$ \d_{\bar\t} x_0(\t) = 0 ,$
\par
  ii) $\t$ must be a "left" solution of the retarded
 time equation - $\t_L$, corresponding to an intersection of the "left" null
plane with the world line $x_0(\t)$.
\par
And also we obtain
\begin{equation}
K \l_1 = {\dot x_0}^\m e^1_\m , \qquad K \l_2 = {\dot x_0}^\m
(e^3_\m - \Y e^1_\m), \label{(1.46} \end{equation}
\begin{equation}
P = {\dot x_0}^\m e^3_\m , \qquad P_{\Y}=\partial _{\Y} P =
{\dot x_0}^\m e^1_\m , \qquad  Y,_3 = - Z P_{\Y}/P.
\label{1.47} \end{equation}
\par
\medskip
4. The retarded time construction gives a complex world line as
complex source of the Kerr geometry. Really it forms a world sheet
in complex Minkowski space with the Euclidean parametrization
 $\tau = t+i\sigma, \qquad \bar \tau =t - i\sigma, $
or hyperbolic string.
\par
The hyperbolic strings can not be open, because it is
impossible to satisfy the boundary  conditions for the real and
imaginary parts of the open string simultaneously. However, one can
form the closed complex hyperbolic strings with an orbifold-like
world-sheet.
\par
  One can consider Euclidean strings in complex Minkowski space $CM^4$
as  a complex objects with a Hermitian Lagrangian,
\begin{equation} L = - T \eta_{i\bar{j}} ({\partial}_{\tau} X^{i}
{\partial}_ {\bar \tau } {\bar X}^{\bar{j}} + {\partial} _{\bar \tau }
X^{i} {\partial} _{\tau}
{\bar X }^{\bar{j }}),\label{eq:(3.23)}\end{equation}
where T is a parameter of string tension.
The  left and right modes are not necessarily complex conjugates of
each other.
An oscillator  expansion of the hyperbolic strings
contains the hyperbolic basis functions, which are not orthogonal
over the string length. It is one of the main obstacles to get the usual
string theory.
To form  the orbifold-like world-sheet it has to be
modified to admit the twisting
boundary conditions for the imaginary part of the complex string.
The interval $\Sigma =[-a, a]$ of the string parameter $\sigma$
has to be doubled. Two copies of the interval
  $\sigma _{+} \in
\Sigma _{+} = [-a, a]$ and  $\sigma _{-} \in
\Sigma _{-} = [-a, a]$
are joined  to  form  an (oriented)
circle ${\bf S}^{1}= \Sigma _{+} \cup \Sigma _{-}.$
which may  be
parametrized  by  a  periodic coordinate $\theta  \sim \theta + 2\pi $
and represents  a covering space for the original interval $\Sigma $ .
The original string is then parametrized by $0 \leq \theta \leq \pi$,
  and second sheet $\pi\leq \theta \leq 2\pi$
covers the string in the opposite direction.
Thus there is an extra symmetry  that "turns over" the world-sheet.
\par
Group $ G ={\bf Z}_2; \quad  g\in G,$ acts on the circle $g^{2}= 1$
  and induces the transformation $ g \sigma _{+} = - \sigma _{-}
;\qquad g \sigma _{-}= - \sigma _{+};$  $ g \tau_{+}= \bar{\tau}_{-}
  ;\qquad g \tau_{-}= \bar{\tau}_{+}. $
  The string modes $X_L (\tau_L)$ and $X_R (\bar \tau _R)$ can
be extended on the circle for $\pi \leq \theta \leq 2\pi $.
\begin{equation} X_L (g\tau_+) = X_R(\bar \tau _+); \qquad
X_R (g\bar \tau _+) = X_L(\tau_-) \label{eq:ext}
\end{equation}
The group G is the symmetry group of the theory and therefore acts on
  the string variables
\begin{equation} \tilde{g}X_L(\tau_\pm) =X_L (g\tau _\pm) = X_R
(\bar \tau _\pm),
\qquad \tilde{g}X_R(\bar \tau _\pm) =X_R (g\bar \tau  _\pm) = X_L (\tau_\pm).
  \label{ext} \end{equation}
 One can see that the group acts by interchanging  the right and left
modes.
\par
The  time coordinate $t$ may be transformed by projection on
the circle $S^1$ and the world-sheet forms the torus ${\bf T}^{2} =
{\bf S}^{1} \times  {\bf S}^{1}$ .  The action of the time reversing
element $h, $ of ${\bf Z}_{2} $ is $h t = - t, \quad h^2 =1$. The
 elements $g$ and $h$  commute with each other.
The quotient space
${\bf T}/{\bf Z}_{2}$  is the orbifold [25].
\footnote{ P. Horava informed us recently that this construction
is a very special case of the "world-sheet orbifolds" that he
developed in [21].}
\par
The Hilbert space of the string solutions H on the orbifold is
decomposed into two subspaces: the $H_{+}$  space of the even functions
\begin{equation} X_{ev} (t,\sigma ) =
( X + \tilde g X)/2 , \qquad \tilde g X_{ev} =
X_{ev}, \end{equation}
and the $H_{-}$  space of the odd functions
\begin{equation} X_{odd} (t,\sigma ) =
(X - \tilde g X)/2;\qquad\tilde g X_{odd} = - X_{odd}.\end{equation}
As a result the even solutions are closed on the orbifold
and the odd solutions form the closed strings with twisted boundary
conditions. A general complex string solution on the orbifold may be
 represented as the sum of the even and odd parts $ X(t,\sigma ) =
 X_{ev} + X_{odd} .$
\par
{}From the  string constraints  it follows  that there exists
one peculiar class of solutions with $X_R = 0$ .
Such solutions contain only left modes and correspond to analytical
world-sheets.
\par
        It has the consequence that the complex Kerr source
satisfies the hyperbolic string equations on the orbifold.
The total momentum of the hyperbolic string is
\begin{equation} P^{i} = \int^a_{-a} {\cal P}_t^{i} d \sigma =
(T/2) \int^a_{-a}
({\partial}_\tau + {\partial}_{\bar \tau }){ X }^{i} d \sigma.
\label{mom} \end{equation}  \noindent
For the straight analitical line  it gives
the momentum $  P^{i} = 0,  i=1,2,3 $
and mass $$ m = P^0 = T a. $$
Thus the  complex string source adds the extra
relation $m=Ta$ to the basic relation for the parameters of the Kerr
solution $ J = m a $ and leads to the Regge dependence of angular momentum
\begin{equation} J = (1/T) m^2. \label{Red} \end{equation}
\par
\medskip
5. The orbifold structure of world sheet is connected with an orbifold
structure of the real and complex Kerr geometry. The "right" and "left"
planes change their places under the action of the
gh -transformation $ (v -v_o) \longleftrightarrow (u-u_o)$  and a
 redefinition $\tilde{Y} = -1/Y$. Simultaneously
 the form (\ref{1.7}) changes the sign and
the retarded and advanced folds of the complex light cone change their
places, that can be seen from alternative splitting of the complex light
cone \begin{equation} t-\tau = \pm \tilde{ r} ;\qquad
gh \tilde{ r} = - \tilde {r}, \label{Split}\end{equation}
where $ (\tilde{r})^2= (x-x_o)^2+(y-y_o)^2 +(z-z_o)^2 $.
When $t$ is real the transformation $ g\tau =\bar \tau $
acts on  $\tilde{r} $ as complex conjugation.
\par
Thus, the gh-transformations act simultaneously on the world-sheet and on
the target space and introduce an equivalence between the left and right
planes and between the positive and negative folds of light cone.
\par
  To obtain
the real cut of the complex Kerr - Schild space one can use a null
vector ${\cal M }^{i} = \psi  \sigma ^{i}\tilde{\psi }$
of the complex light cone, linking   points  of  the  complex
world line $X_{o}(\tau)$ with  points of the real slice.
If a cone has real slice, then
$X_{o}+ {\cal M }$ is a point on the real slice and
one can also add to  $\cal M  $ a real vector directed along the principal
 null congruence $ r e^3 $.
The null vectors $ e^3 $ and $ {\cal M}$ belong to the
"left" null plane.
\par
On the other hand for every real point of the Kerr-Schild space $X^i$
one can find both
the "left" retarded time $\tau_L$ and a corresponding point of the complex
 world-line $X_o(\tau_L)$ defined by intersection of the world-line
with the "left" null plane (\ref{L}) and this plane is spanned by the null
vectors ${\cal M} ^i $ and $e^3$.
The "right" plane (\ref{R}) contains the vector
${\cal M} $, but it also defines the second null congruence
 $l^i=\tilde{\bar \psi } \sigma^i\tilde{\psi} $
through the equation $\tilde{Y}= \tilde{\psi^1} /\tilde{\psi^0}$.
\par
On the real slice of the Kerr geometry we have  the orbifold
structure in the form of well-known twosheetedness of the Kerr space.
Let us consider the coordinate surface $ \phi = const $ which is generated
by the real rays (space-like projection) of the Kerr congruence.
\footnote{The generating ray, corresponding to the value of
$\theta$, may be obtained from
the Z-axis by means of the shift along the X-axis at $ a sin(\theta) $ and
rotation on the angle $\theta$ in the plane orthogonal to the shift.}
There are two folds of this surface corresponding to negative and
positive values of $r$. One can see that they cover the real
Kerr-Schild space twice that corresponds two principal null
congruencies for the type D metrics. Since $ hr = -r$ these folds also
change their positions by the gh-transformation and changing of
the advanced and retarded folds of the light cone.
An introduction of the orbifold by fixing of the gh-equivalence leads to a
truncation of the negative sheet of the Kerr space as a mirror image of the
positive one in a disk-like mirror. Therefore, the introducing of orbifold
corresponds the disk-like version for the real Kerr source.
\par
\bigskip
6.In the "realistic" realization of source the string-like
singularity has to be considered as some kind of idealization. After
the truncation of the negative sheet of space a disk-like matter source
has to be introduced, which represents a relativistic rotating disk-like
shell (membrane) spanned by the singular ring.
The "negative" sheet of metric may be considered as an image in the
superconducting mirror. The singular ring turns into a ring-like vortex line
of superconductor. In the model of microgeon with the Kerr metric [3,4],
this singular ring was used as a waveguide for  wave excitations. Similar
idea has been used recently in dilaton gravity to construct the black
string solutions with traveling waves [22].
Recently, in the theory with axion and dilaton,
the exact generalizations of the Kerr solution has been obtained [23],
and there has been also increasing interest in
the extremal black hole as a model of elementary particle [24].
However, in the case of massless dilaton, the charge/mass ratio for the
extremal black holes is too far from that of elementary particles [25].
Nevertheless, the stringy corrections with a massive
dilaton may be useful to get a more realistic model of the Kerr metric
source.
\par
\bigskip
{\bf Note added 2th March 1994:}
\medskip
Recently we have found an important additional argument in favor of the
old proposal of Professor Dmitri Ivanenko and me that the Kerr
singular ring plays the role of string [4].
The Kerr solution in axion-dilaton gravity may be considered as
a fundamental soliton-like solution of the heterotic string theory [23].
Recent analysis of this solution for the case of the big angular momentum
$\mid a \mid \gg m$
shows that  axidilaton field concentrates near the Kerr singular ring.
It was obtained also that the limiting form of this solution near the
singular ring coincides with the solution obtained by Sen [23] for
the field around a fundamental heterotic string. Thus, the Kerr singular
ring becomes an interpretation as a heterotic string. It is worth
mentioning that the axidilaton field does not remove the  Kerr singularity
and the problem of real source of the Kerr solution retains its importance.
\par
\pagebreak
REFERENCES:
\par
\noindent 1. B. Carter,Phys. Rev. {\bf 174}(1968) 1559.
\par
\noindent 2. W. Israel, Phys rev ${\bf D2} (1970) 641.$
\par
\noindent 3. A.Ya. Burinskii, Sov. Phys. JETP {\bf 39}(1974) 193.
\par
\noindent 4. D. Ivanenko and A.Ya. Burinskii, Izvestiya Vuzov Fiz.{\bf 5}
(1975) 135 , {\bf 7} (1978)113 (in Russian, transl.: Sov. Phys. J. (USA)).
\par
\noindent 5. A.Ya. Burinskii, in: Problem of theory of gravitation and
elementary particles,{\bf11}(1980), Moscow, Atomizdat, (in Russian).
\par A.Ya. Burinskii, Izvestiya Vuzov Fiz.{\bf 5}(1988)82 (in Russian,
transl.: Sov. Phys. J. (USA)).
\noindent 6. A.Ya. Burinskii, Phys. Lett.{\bf B 216} (1989) 123.
\par
\noindent 7. C.A. L\'opez, Phys. Rev. ${\bf D30} (1984) 313.$
\par
\par
\noindent 8. I.Tiomno, Phys. Rev.{\bf D7}(1973) 992.
\par
\noindent 9. H.B.Nielsen and P. Olesen, Nucl. Phys.,${\bf B61}(1973)45.$
\par
 E. Witten, Nucl.Phys.,${\bf B249}(1985)557.$
\par
\noindent 10. A. Burinskii , Phys. Lett. {\bf A} 185 (1994) 441-445.
\par
 A.Burinskii, String-like Structures in Complex Kerr Geometry,
in: Relativity Today, Proceedings of the Fourth Hungarian Relativity
Workshop, Edited by R.P. Kerr and Z. Perj\'es,  Academiai Kiado, Budapest
1994 p.149, preprint gr-qc/9303003
\par
\noindent 11.
 A.Burinskii, Proceedings of the VII All-union
Gravitational Conference. 18-20 October 1988 p.255-257, USSR, Yerevan.
\par
\noindent 12.  E.T. Whittacker and G.N. Watson, A Course of Modern Analysis,
Cambrige Univ. Press London/New York,p.400, 1969 .
\par
\noindent 13.  E.T. Newman, J.Math.Phys.,{\bf 14}(1973)102.
\par R.W. Lind, E.T. Newman, J. Math. Phys.,{\bf 15}(1974)1103.
\par
\noindent 14.
H.Ooguri, C. Vafa, Nucl. Phys.{\bf B 361}, 469
(1991)
\par
\noindent 15. E. Witten, Phys.Rev.{\bf D44}, 314 (1991)
\par
\noindent 16. P.Horava, Phys.Lett.{\bf B278} 101 (1992)
\par
\noindent 17. G.C. Debney, R.P. Kerr, A.Schild, J.Math.Phys.,{\bf 10}(1969)
1842.
\par
\noindent 18. A. Burinskii, R. Kerr and Z. Per\'jes,  Nonstationary
 Kerr Congruences, preprint gr-qc/9501012
\par
\noindent 19. R. Penrose,J. Math. Phys.,{\bf 8}(1967) 345.
\par
\noindent 20. R.P. Kerr, W.B. Willson, Gen. Rel. Grav.,{\bf 10}(1979)273.
\par
\noindent 21. L. Dixon,J.A.Harvey, C.Vafa, E.Witten, Nucl. Phys.
{\bf B261}(1985)678.
\par
\noindent 22. D. Garfinkel, Black String Traveling Waves, gr-qc/9209002
\par
\noindent 23. A. Sen,  Black Holes and Solitons in String Theory,
hep-th/9210050.
\par
 D.V. Gal'tsov, O.V. Kechkin,
Ehlers--Harrison--Tipe Transformations in Dilaton-Axion Gravity,
hep-th/9407155
\par
\noindent 24. J. Preskill, P.Schwarz, A. Shapere, S.Trivedi and F.Wilczek,
Mod. Phys. Lett. {\bf A6}, 2353 (1991);
\par
\noindent 25. J.Horne and G. Horowitz, Black Holes Coupled to a Massive
Dilaton, hep-th/9210012
\end{document}